\documentclass[aps,pra,twocolumn]{revtex4}
\usepackage{graphicx}
\usepackage{dcolumn}
\usepackage{amsmath}
\usepackage{amssymb}
\def\erf{{\rm erf}}
\def\rv{{\bf r}}

\def\Rv{{\bf R}}

\def\beq{\begin{equation}}
\def\eeq{\end{equation}}
\def\vks{v_{\rm KS}}

\begin{document}
\title{Simple model for the spherically- and system-averaged
pair density: \\
Results for two-electron atoms}
\author{Paola Gori-Giorgi and Andreas Savin}
\affiliation{Laboratoire de Chimie Th\'eorique, CNRS,
Universit\'e Pierre et Marie Curie, 4 Place Jussieu,
F-75252 Paris, France}
\date{\today}
\begin{abstract}
As shown by Overhauser and others, accurate pair densities for
the uniform electron gas may be found by solving a two-electron
scattering problem with an effective screened electron-electron
repulsion. In this work we explore the
extension of this approach to nonuniform systems, and we discuss
its potential for density functional theory.
For the spherically- and system-averaged 
pair density of two-electron atoms
 we obtain very accurate short-range properties, including,
for nuclear charge $Z\ge 2$,
``on-top'' values (zero electron-electron distance) 
essentially indistinguishable from those coming from
precise variational wavefunctions. By means of
a nonlinear adiabatic connection that separates
long- and short-range effects, we also obtain Kohn-Sham correlation
energies whose error is less than 4~mHartree, again for $Z\ge 2$,
and short-range-only correlation energies whose accuracy is one order
of magnitude better.
\end{abstract}

\maketitle
\section{Introduction and summary of results}
Density Functional Theory (DFT)~\cite{kohnnobel,science,FNM} 
is nowadays the most
widely used method for electronic structure calculations,
in both condensed matter physics and quantum chemistry, thanks
to a combination of low computational cost and reasonable
accuracy. 
\par
In the application of this theory within the Kohn-Sham (KS)
formalism~\cite{kohnsham}, one deals with a model system (the KS system)
of $N$ noninteracting electrons in a local
potential $\vks(\rv)$ that forces them to yield the same 
density $n(\rv)$ of the physical system. The energy of the
physical system is then obtained from that of the KS system
via a functional of the density, whose only term not explicitly
known is the exchange-correlation energy $E_{xc}[n]$.
Correspondingly, in the local potential $\vks(\rv)$
there is an unknown term,
$v_{xc}(\rv)=\delta E_{xc}[n]/\delta n(\rv)$.
\par
The success of KS DFT is mostly due to the fact that even simple physical
approximations of $E_{xc}[n]$,  like the local density approximation
(LDA)~\cite{kohnsham}, already give acceptable results
for many purposes. This spurred 
fundamental research in the field, and led
to a wealth of more and more sophisticated exchange-correlation 
functionals~\cite{science,FNM,jacob}, and to the development
of different approaches to DFT~\cite{adiabatic,sahni}.  
\par

Recently, in the search for accurate $E_{xc}[n]$, the focus of a large part of
the DFT community has shifted from seeking
{\em explicit} functionals of the density like the generalized gradient 
approximations (GGA)~\cite{GGA}, to {\em implicit}
functionals, tipically  using the Kohn-Sham orbital 
kinetic energy density~\cite{TPSS} or the Kohn-Sham 
orbitals (see, e.g.,~\cite{FNM,newjohn,mike}).
The so-called ``third generation'' of exchange-correlation 
functionals is based on the exact exchange 
of the noninteracting (KS) system, simply obtained
by putting in the formal expression for the Hartee-Fock exchange the
Kohn-Sham orbitals $\varphi_{i\sigma}(\rv)$. Such expression
corresponds to an implicit functional of the density, 
$E_x[n]=E_x[\{\varphi_{i\sigma}[n]\}]$.
The local potential $v_x(\rv)=\delta E_x[n]/\delta n(\rv)$ that generates
the orbitals $\varphi_{i\sigma}(\rv)[n]$
can be obtained via the optimized effective 
potential method (OEP)~\cite{oep}.
\par
\begin{figure*}
\includegraphics[width=7.2cm]{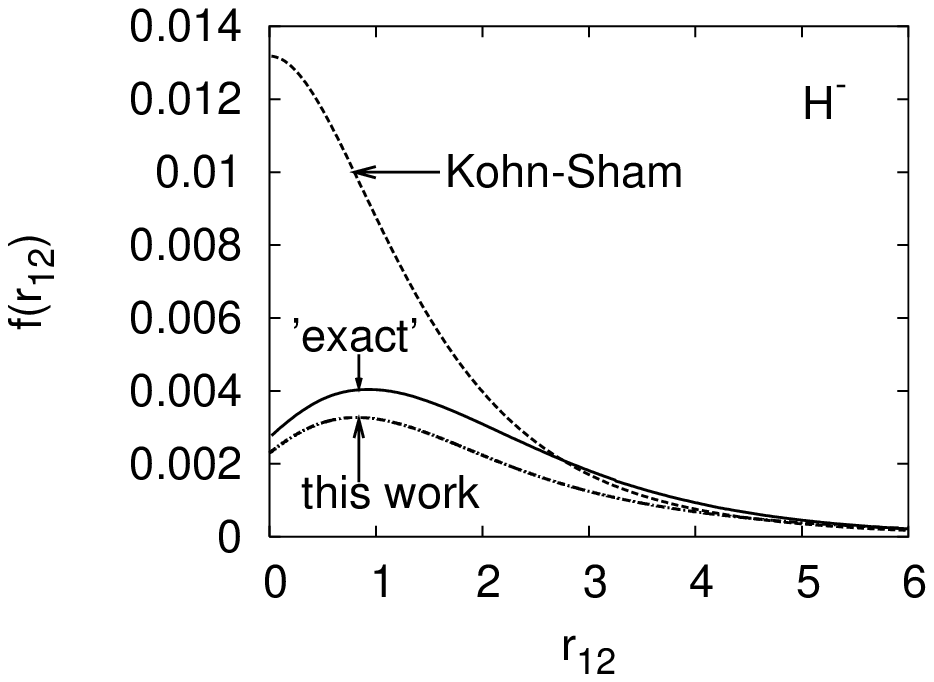} 
\includegraphics[width=7.2cm]{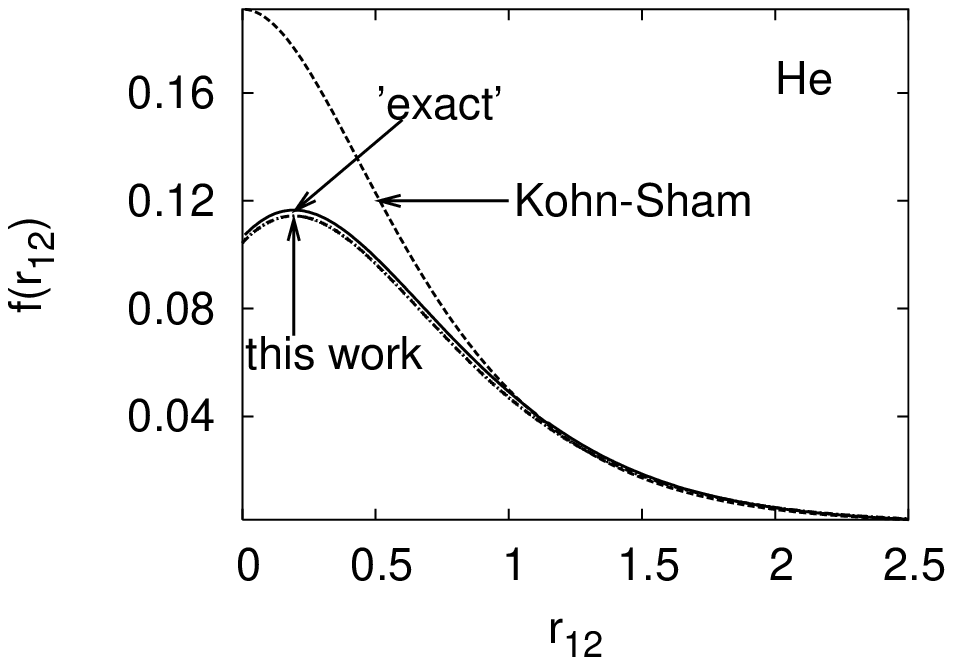} 
\includegraphics[width=7.2cm]{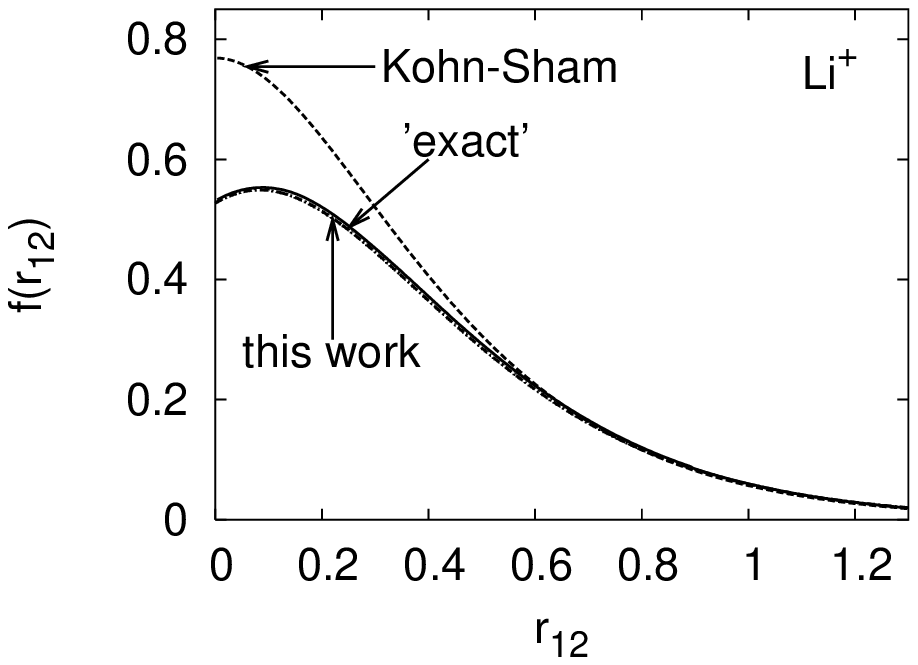} 
\includegraphics[width=7.2cm]{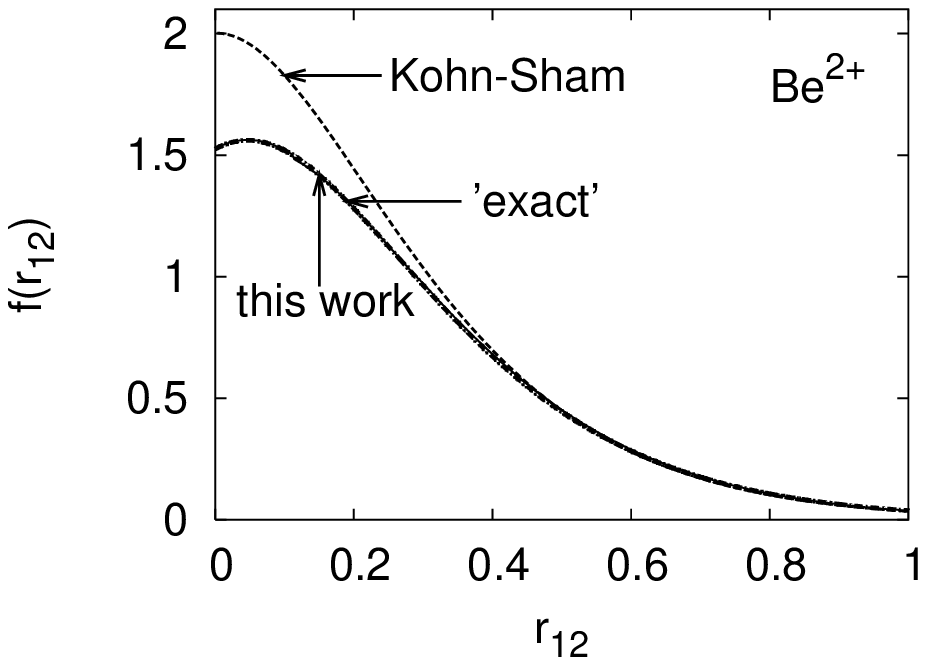} 
\caption{Spherically and system-averaged pair densities
for two-electron atoms: 'exact' results~\cite{morgan} are compared with
the values obtained for the Kohn-Sham system and with
the present approach, which is designed to get realistic
$f(r_{12})$ starting from the Kohn-Sham ones.}
\label{fig_all}
\end{figure*}
In this broad context, sketchily summarized here, 
we propose a simplified method to build the ``bridge''
between the physical and the KS system, or, more generally,
with a reference model system of partially interacting electrons. 
We focus on a quantity which is known to play a crucial
role in DFT and has an intuitive physical meaning,
the spherically and system-averaged electronic pair density
$f(r_{12})$ (also known in chemistry as 
spherical average of the intracule 
density, see e.g.~\cite{Coulson,Cioslowski1,Cioslowski2,ugalde1}, 
and especially~\cite{davidson,coleman}). Given the 
spin-resolved diagonal of the two-body reduced density matrix,
\beq
\gamma^{(2)}_{\sigma_1\sigma_2}(\rv_1,\rv_2)
=\sum_{\sigma_3...\sigma_N} 
\int|\Psi(\rv_1\sigma_1,...,\rv_N\sigma_N)|^2 d\rv_3...d\rv_N,
\label{eq_gamma2}
\eeq
we define the spin-summed
pair density $n_2(\rv_1,\rv_2)$,
\beq
n_2(\rv_1,\rv_2)=\frac{N(N-1)}{2}\sum_{\sigma_1\sigma_2}
\gamma^{(2)}_{\sigma_1\sigma_2}(\rv_1,\rv_2),
\eeq
and we integrate it over all variables but $r_{12}=|\rv_2-\rv_1|$
by switching, e.g.,  to center-of-mass
coordinates, $\Rv=\tfrac{1}{2}(\rv_1+\rv_2)$,
$\rv_{12}=\rv_2-\rv_1$, 
\beq
f(r_{12})=\int d\Rv \,\frac{d\Omega_{{\rv}_{12}}}{4\pi}\,
n_2\left(\Rv-\frac{\rv_{12}}{2},\Rv+\frac{\rv_{12}}{2}\right).
\label{eq_def}
\eeq
The function $f(r_{12})$ times the volume element $4 \pi r_{12}^2\,dr_{12}$ is
proportional to the probability density for the particle-particle 
distance in a system
of $N$ electrons in the state $\Psi$, and is normalized to
the number of electron pairs, $N(N-1)/2$ . This
quantity fully determines the expectation
value of the electronic Coulomb repulsion (in Hartree atomic units
used throughout), 
\beq
\langle V_{ee}\rangle
\equiv \langle\Psi|V_{ee}|\Psi\rangle = \int_0^{\infty} 
4 \pi\, r_{12}^2\, \frac{f(r_{12})}{r_{12}}\,d r_{12},
\label{eq_Vee}
\eeq
and is a measurable quantity, being essentially the Fourier transform
of the electronic static structure factor~\cite{sacchetti}.
By construction, the one-electron density $n(\rv)$ is the same
in the KS and in the physical system, whereas $f(r_{12})$ will be
different in the two cases, as shown, e.g., in Fig.~\ref{fig_all} for
some two-electron atoms. In the physical system $f(r_{12})$ has 
a much lower ``on-top'' value $f(r_{12}=0)$ than in the KS
system, and it has a
cusp~\cite{cusp}, as expected from the fact that
the electrons repel each other via the Coulomb interaction. 
Roughly speaking, in the classic DFT approach to correlation,
the difference in energy arising when
we evaluate the r.h.s. of Eq.~(\ref{eq_Vee}) with the two
$f(r_{12})$, the physical and the KS, is what one tries to describe
with a {\em universal} functional of
the density~\cite{nota}.
Here we follow a different approach: we try to build
realistic $f(r_{12})$ from a set of simple radial equations, to be
solved for {\em each} system, and eventually coupled to a DFT calculation. 

Our approach is inspired by
the seminal work of Overhauser~\cite{Ov} and its subsequent 
extension~\cite{GP1}, in which the function $f(r_{12})$ for the uniform
electron gas is obtained from a set of
geminals, solutions of a radial Schr\"odinger equation with an effective
electron-electron ($e$-$e$) potential. 
Simple approximations for such effective $e$-$e$ potential
give indeed accurate results at all relevant 
densities~\cite{GP1,DPAT1,CGPenagy2}. Here we try to generalize this approach
to systems of nonuniform density to get accurate $f(r_{12})$. The main goal of 
the present work is understanding whether the method is promising, and
whether it is worth developing and refining it. 
To this purpose, we define
the formalism (Sec.~\ref{sec_formalism}), 
we give a physically-motivated  prescription 
for the effective $e$-$e$ potential (Sec.~\ref{sec_potential}), 
and we test it on the simple but not trivial case
of two-electron  atoms (Sec.~\ref{sec_results}). 
The prescription for the effective $e$-$e$ potential
used here is not very sophisticated. Improvements along the lines
of what has been done for the uniform
electron gas~\cite{GP1,DPAT1,CGPenagy2} will be the subject of future work. 
Yet, even at this simple first stage of the theory we already obtain
rather accurate results, especially for the
short-range part of $f(r_{12})$ (see Fig.~\ref{fig_all} and 
Table~\ref{tab_resume}). In Sec.~\ref{sec_adiabatic} we show that with the 
present approach we can
also recover the difference in kinetic energy between the physical
and the KS system. Finally, Sec.~\ref{sec_last} is devoted to
conclusions, perspectives and open questions.

\section{Formalism}
\label{sec_formalism}
In addition to the work on the ``Overhauser
model''~\cite{Ov,GP1,DPAT1},
the approach described here takes advantage of inspiring
papers on the possibility of constructing a pair-density functional
theory~\cite{ziesche1,gonis,agnesnagy,furche}, and
a local-density-of-states functional theory~\cite{soler}.
\par
Our starting point is a constrained search over
$\frac{1}{2}N(N-1)$ ``effective'' orthonormal geminals $\psi_i(r_{12})$ that
minimize the electron-electron relative kinetic energy 
$T_{12}=-\nabla^2_{r_{12}}$ (the reduced mass for the relative motion is 1/2)
and yield the exact $f$, $\sum_i |\psi_i(r_{12})|^2=
f(r_{12})$,
\beq
\min_{\{\psi_i\}\to f} \sum_i \langle \psi_i |
-\nabla^2_{r_{12}}| \psi_i \rangle,
\label{eq_geminals}
\eeq
thus leading to a set of radial equations formally similar
to the KS ones,
\begin{eqnarray}
& & [-\nabla^2_{r_{12}}+v_{\rm eff}(r_{12})] \psi_i(r_{12})  =  \epsilon_i\,
\psi_i(r_{12}) 
\label{eq_eff1} \\ 
& & \sum_{i=1}^{N(N-1)/2} |\psi_i(r_{12})|^2  =  f(r_{12}).
\label{eq_eff2}
\end{eqnarray}
These equations imply that an expansion in spherical harmonics
of $f(r_{12})$ has been done, so that the kinetic energy
operator also contains the usual $\ell(\ell+1)/r^2$ term. 
To fully define these equations we need a rule for the
occupancy of the effective geminals.
In analogy with what has been done for the uniform electron 
gas~\cite{GP1,DPAT1}, we can assign spin degeneracy 1 to
even-angular-momentum states (singlet) 
and spin degeneracy 3 to odd-angular-momentum
states (triplet), up to $N(N-1)/2$ occupied states. 
More generally, for open shell systems it could be better
to develop the formalism for the spin-resolved quantities, starting
from Eq.~(\ref{eq_gamma2}). This will be investigated in future work.

The effective electron-electron potential $v_{\rm eff}(r_{12})$ 
of Eq.~(\ref{eq_eff1}) is the Lagrange parameter for $f(r_{12})$, and is
a functional of $f$ itself and of the electron-nucleus external
potential $V_{ne}$ (or, equivalently, of the density $n(\rv)$).
To see this, we can rewrite our Eqs.~(\ref{eq_eff1})-(\ref{eq_eff2})
in terms of a minimization of the total energy in two steps, using
the constrained search formalism~\cite{levy,lieb} for the ground
state energy $E=\min_\Psi \langle \Psi| T + V_{ee}+V_{ne}|\Psi\rangle$,
\begin{eqnarray}
E & = & \min_f \min_{\Psi \to f} \Big\{\min_{\{\psi_i\}\to f} \sum_i
\langle \psi_i | -\nabla^2_{r_{12}}|\psi_i\rangle + 
 \int  
\frac{f}{r_{12}}d \rv_{12} \nonumber \\
& & + \langle \Psi| T + 
V_{ne}|\Psi\rangle  -\min_{\{\psi_i\}\to f} \sum_i
\langle \psi_i | -\nabla^2_{r_{12}}|\psi_i\rangle\Big\}.
\label{eq_step1}
\end{eqnarray}
Defining the kinetic and external-potential functional as
\begin{eqnarray}
  F_{\rm KE}[f;V_{ne}] =   \nonumber \\
\min_{\Psi \to f}\langle \Psi| T + 
V_{ne}|\Psi\rangle  -\min_{\{\psi_i\}\to f} \sum_i
\langle \psi_i | -\nabla^2_{r_{12}}|\psi_i\rangle,
\end{eqnarray}
we can rewrite
\begin{eqnarray}
E & = & \min_f\Big\{\min_{\{\psi_i\}\to f} \sum_i
\langle \psi_i | -\nabla^2_{r_{12}}|\psi_i\rangle + 
 \int  
\frac{f}{r_{12}}d \rv_{12} \nonumber \\
& & +F_{\rm KE}[f;V_{ne}]\Big\}.
\end{eqnarray}
Searching this minimum by directly varying 
the $\psi_i$ (with given, fixed, $V_{ne}$) leads to
Eqs.~(\ref{eq_eff1})-(\ref{eq_eff2}) with the identification
\beq
v_{\rm eff}(r_{12})=\frac{1}{r_{12}}+\frac{\delta F_{\rm KE}[f;V_{ne}]}
{\delta f(r_{12})}.
\label{eq_vefffunz}
\eeq
Thus, in principle we could recover the whole ground-state energy
via the (unknown) system-dependent functional $F_{\rm KE}[f;V_{ne}]$. 
In practice, it seems much
more feasible to combine Eqs.~(\ref{eq_eff1})-(\ref{eq_eff2})
with a DFT calculation, that yields the complementary information
(the density, and thus $\langle\Psi|V_{ne}|\Psi\rangle$). 
The steps of Eqs.~(\ref{eq_step1})-(\ref{eq_vefffunz}) can  be repeated
for arbitrary electron-electron interaction and external one-body
potential. In particular, we can set $V_{ee}^{\lambda}=\lambda V_{ee}$ and
$V_{ne}=V^{\lambda}$, where
$V^{\lambda}$ is an external potential that keeps the
density equal to the one of the physical system. One could thus
obtain  $f^{\lambda}$ at each coupling strength $\lambda$ between
0 and 1 from Eqs.~(\ref{eq_eff1})-(\ref{eq_eff2}) with a suitable 
$v_{\rm eff}^{\lambda}$. The correlation energy of KS theory
is then simply given by~\cite{gunnarsson,adiabatic,wang}
\beq
E_c[n]=\int_0^1 d\lambda\int d\rv_{12}\, \frac{f^{\lambda}(r_{12})-
f^{\lambda=0}(r_{12})}{r_{12}}.
\label{eq_adialinear}
\eeq
Alternatively, this procedure (usually called adiabatic 
connection~\cite{wang}) can
be performed along a nonlinear path, e.g., by 
setting~\cite{adiabatic,erf,sav_madrid,julien}
$v_{ee}^\lambda=\erf(\lambda r)/r$, where $\erf(x)$ is the error
function (see Sec.~\ref{sec_adiabatic}).
Eventually, the two sets of equations, KS and~(\ref{eq_eff1})-(\ref{eq_eff2})
plus (\ref{eq_adialinear}),
could be solved together self-consistently. This last issue is discussed in
Sec.~\ref{sec_last}. Notice that if we combine Eqs.~(\ref{eq_eff1}), 
(\ref{eq_eff2}) and (\ref{eq_adialinear}) with a DFT calculation, 
we only need to approximate the potential $v_{\rm eff}^\lambda(r_{12})$ 
and not the whole functional $F_{\rm KE}$ since
the remaining information is provided by DFT.

It is also worth to stress at this point that there is no wavefunction 
behind our Eqs.~(\ref{eq_eff1})-(\ref{eq_eff2}):
the effective geminals $\psi_i$ are defined via Eq.~(\ref{eq_geminals}), and
by specifying their occupancy (e.g., triplet and singlet). A bosonic
version of the theory, in which only one geminal (proportional
to $\sqrt{f(r_{12})}$) is occupied can also be
considered~\cite{furche,tosi2}. In this work we only focus on
two-electron systems for which the two choices are
equivalent. A careful comparison of performances of the ``fermion-like'' and
of the ``boson-like'' occupancy in the uniform electron gas is the subject
of current investigations~\cite{pisani_new}.

As for KS DFT,
the formalism just described can be useful only if simple approximations
for $v_{\rm eff}(r_{12})$ yield accurate results. This is what we
start to check in the rest of this paper.
First, we construct a physically-motivated $v_{\rm eff}$ for
two-electron atoms
for the fully-interacting system, and we compare our results
with ``exact'' ones~\cite{morgan}. Then, we 
generalize our construction to build $v_{\rm eff}$ along the 
adiabatic connection, and
we calculate the KS correlation energy. 


\section{Effective electron-electron potential:
the Overhauser model} 
\label{sec_potential}
For the interacting electron gas of uniform density $n$,
Overhauser~\cite{Ov} proposed a simple and reasonable
effective potential $v_{\rm eff}(r_{12})$:
he took the
sphere of volume $n^{-1}$ around a given electron as the boundary 
within which the other electrons are excluded, due to exchange
and correlation effects.
In the standard uniform-electron-gas
model, a rigid positively-charged background maintains the
electrical neutrality. Thus the exclusion region (or ``hole'') around
a given electron, modeled with
a sphere of radius $r_s=(4 \pi n/3)^{-1/3}$, uncovers the background
of positive charge, leading to an
effective screened Coulomb potential with screening length $r_s$,
\beq
v_{\rm eff}^{Ov}(r_{12};r_s)=\frac{1}{r_{12}}-
\int_{|\rv|\le r_s} \frac{n}{|\rv-\rv_{12}|}\,d\rv,
\eeq
equal to
\begin{eqnarray}
  v_{\rm eff}^{Ov}(r_{12};r_s)    =  & \frac{1}{r_{12}}
+\frac{r_{12}^2}{2r_s^3}-\frac{3}{2 r_s}
  \qquad  &  r_{12}\le r_s \nonumber  \\
 v_{\rm eff}^{Ov}(r_{12};r_s)    =  & 0    & r_{12}>r_s. \label{eq_potOv}
\end{eqnarray} 
Equations~(\ref{eq_eff1})-(\ref{eq_eff2}),
combined with the Overhauser effective potential of Eq.~(\ref{eq_potOv})
gave extremely accurate results
for the short-range part ($r_{12}\le r_s$) of the function 
$f(r_{12})$ in the uniform electron
gas at all relevant densities~\cite{GP1}. 
A more sophisticated effective potential, based on 
a self-consistent Hartree
approximation, extended such accuracy to the long-range part
of $f(r_{12})$ at metallic densities~\cite{DPAT1}. 
Other approximate $v_{\rm eff}(r_{12})$
for the uniform electron gas have also been proposed~\cite{CGPenagy2},
and exact properties have been derived~\cite{ziesche2}.

To produce realistic $f(r_{12})$ for nonuniform
systems from Eqs.~(\ref{eq_eff1})-(\ref{eq_eff2}), here
we generalize the original idea of Overhauser~\cite{Ov,GP1} 
to two-electron atoms, and show
that it gives rather accurate results, especially for the short-range
part of $f(r_{12})$. We start from the effective potential 
$v_{\rm eff}^{(0)}(r_{12})$ 
that generates $f_{\rm KS}(r_{12})$, the spherically-
and system averaged pair density of 
the Kohn-Sham system. In the special case of a
spin-compensated two-electron
system, the KS wavefunction is simply equal to $\frac{1}{2}
\sqrt{n(\rv_1)}\sqrt{n(\rv_2)}$. 
Because, at this first stage, we are interested in testing
our method as a ``bridge'' between the KS and the real system, here
we use the ``exact'' Kohn-Sham system. 
We thus take  accurate
one-electron densities~\cite{morgan}, and construct 
$f_{\rm KS}(r_{12})$, 
\beq
f_{\rm KS}(r_{12})=
\frac{1}{4}\int  
n\left(\Rv-\frac{\rv_{12}}{2}\right)
n\left(\Rv+\frac{\rv_{12}}{2}\right)\,d\Rv  
\frac{d\Omega_{\rv_{12}}}{4 \pi},
\eeq
and the corresponding ``exact'' potential $v_{\rm eff}^{(0)}(r_{12})$,
that can be calculated
by inverting Eqs.~(\ref{eq_eff1})-(\ref{eq_eff2}),
\beq
v_{\rm eff}^{(0)}=\frac{\nabla^2 \sqrt{f_{\rm KS}}}{\sqrt{f_{\rm KS}}}+
{\rm const.}
\eeq
For systems with more than two electrons, the potential
$v_{\rm eff}^{(0)}$ could be calculated, e.g., with the methods
of Refs.~\cite{francois,parr}.
In practice, it would be much more efficient to 
build approximations also for $v_{\rm eff}^{(0)}$ (see Sec.~\ref{sec_last}).
Examples of functions $f_{\rm KS}$ for nuclear charges 
$Z=1,2,3,4$ are given
in Fig.~\ref{fig_all}: they have a maximum 
at $r_{12}=0$, as expected in
 a system of two
non-interacting electrons with antiparallel spins
in a confining one-body external potential.
When the interaction is turned on, the average distance between the two 
electrons increases, with the constraint that $n(\rv)$ is kept
fixed.
We can thus imagine that, 
with respect to the Kohn-Sham system, in the physical system 
the Coulomb repulsion
between the electrons creates, on average, a screening ``hole'' around
the reference electron of volume $(\overline{n})^{-1}$, where
$\overline{n}$ is an average density (i.e., $n(\rv)$ integrated over
the wavefunction),
\beq
\overline{n}=\frac{1}{N}\int d\rv\,n(\rv)^2.
\label{eq_avn}
\eeq
An approximate $v_{\rm eff}(r_{12})$ could thus be simply constructed as
\beq
v_{\rm eff}(r_{12}) \approx v_{\rm eff}^{(0)}(r_{12})+v_{\rm eff}^{Ov}
(r_{12};\overline{r}_s) 
\label{eq_ourv}
\eeq
with an average $\overline{r}_s$ in $v_{\rm eff}^{Ov}$
of Eq.~(\ref{eq_potOv}),
\beq
\overline{r}_s=\left(\tfrac{4 \pi}{3}\,\overline{n}\right)^{-1/3}.
\label{eq_rsav}
\eeq
The Overhauser-like potential $v_{\rm eff}^{Ov}(r_{12};\overline{r}_s)$
is thus a correlation potential to be added to the one that generates
$f_{\rm KS}$. 
It describes the correlation between pairs of electrons due
to Coulomb interaction, and keeps the information on the one-electron
density in an approximate way, via the average $\overline{n}$ of
Eq.~(\ref{eq_avn}). Of course, for more complicated systems we expect to need a 
more sophisticated construction for $\overline{r}_s$.
\begin{table}
\begin{tabular}{llllll}
\hline\hline
 &  H$^-$ & He & Li$^+$ & Be$^{2+}$ & Ne$^{8+}$ \\
\hline
 $\overline{r}_s$ &  2.1   &  0.86  &  0.54 & 0.39 & 0.15 \\
& & & & \\  
 $f(0)$ & 0.0021 &	0.104 &	0.528	& 1.526 & 32.6\\
``exact'' & 0.0027 & 0.106  & 0.534 & 1.523 & 32.7\\  
LDA &      0.0047 & 0.119 & 0.563 & 1.587 & 33.0\\
& & & & \\  
$r_{12}^{\rm max}$ & 0.835 & 0.193 & 0.083 & 0.0465 & 0.0074\\
``exact''          & 0.927 & 0.194 & 0.083 & 0.0465 & 0.0074\\
& & & & \\
$f(r_{12}^{\rm max})$ & 0.0031	& 0.114	& 0.55	& 1.56 & 32.74\\
``exact''             & 0.0040  & 0.117 & 0.56 & 1.56 & 32.74\\
& & & & \\
$\langle V_{ee} \rangle-\langle V_{ee} \rangle_{KS}$
           &  -0.12 &  -0.097 &
  -0.10 &  -0.10 &  -0.10\\
``exact'' & -0.07 & -0.078 & -0.082 & -0.089 & -0.09\\
\hline\hline  
\end{tabular}
\caption{Our results for the function $f(r_{12})$ for two-electron atoms
(first line for each property) compared
with the corresponding ``exact'' quantities~\cite{morgan}. 
In the first line of the table we report
the average $\overline{r}_s$ as defined by Eqs.~(\ref{eq_avn}) 
and~(\ref{eq_rsav}).
For the ``on-top'' value $f(0)$ we also show the LDA result (with
$f(0)$ for the uniform electron gas from Ref.~\cite{GP1}).
All values are in Hartree atomic units.}
\label{tab_resume}
\end{table}

\section{Results}
\label{sec_results}
We have inserted the potential of Eq.~(\ref{eq_ourv}) into 
Eqs.~(\ref{eq_eff1})-(\ref{eq_eff2}), and solved them for several
two-electron atoms.
Our results are shown in Fig.~\ref{fig_all}
and summarized in Table~\ref{tab_resume}. We see that
the simple effective potential of Eq.~(\ref{eq_ourv}) gives
already reasonable results for $Z=1$ and 2, and that the accuracy of the
results increases with $Z$ (as the system
becomes less and less correlated). The ``on-top'' value $f(0)$ is essentially
exact for $Z\ge 2$, and is much better than the LDA estimate (normally
regarded as accurate) for all $Z$.
This feature is appealing, since the on-top value plays an important
role in DFT~\cite{ontop}, and accurate $f(0)$ are not
easy to obtain from {\it ab initio} methods (see, e.g., 
Ref.~\cite{ugalde2000} and references therein). 
The term $1/r_{12}$ in the effective potential ensures that the calculated
$f(r_{12})$ satisfies the exact cusp condition $f'(0)=f(0)$.
Table~\ref{tab_resume} also shows that the position $r_{12}^{\rm max}$ 
and the height $f(r_{12}^{\rm max})$
of the maximum of $f$ is very well predicted by the present approach.
The presence of this maximum is essentially due 
to the combined effect of the Coulomb repulsion between
the electrons and the confining external potential.

\begin{figure}
\includegraphics[width=6.5cm]{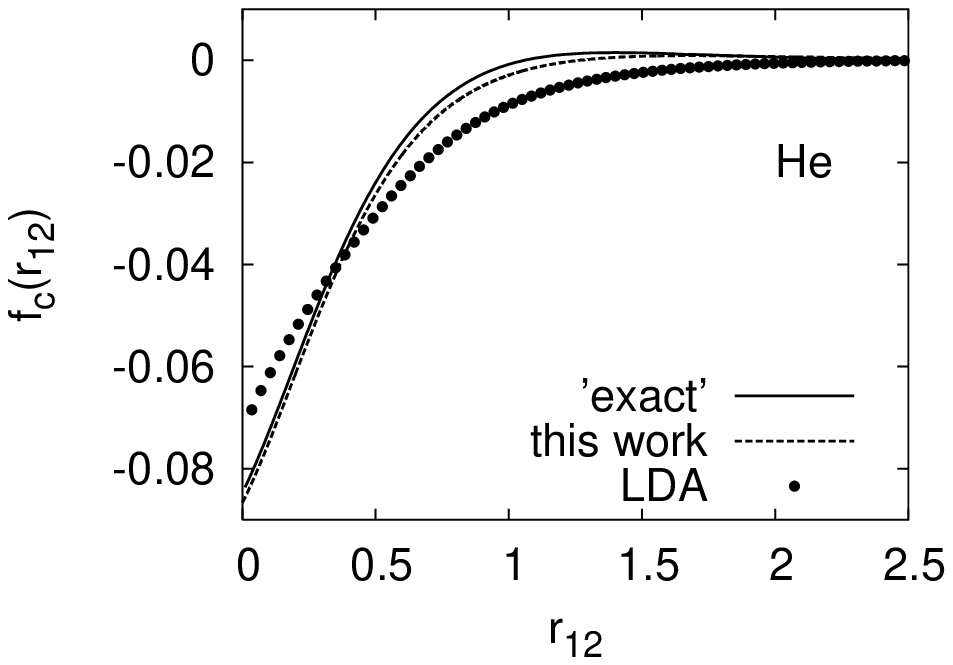} 
\includegraphics[width=6.5cm]{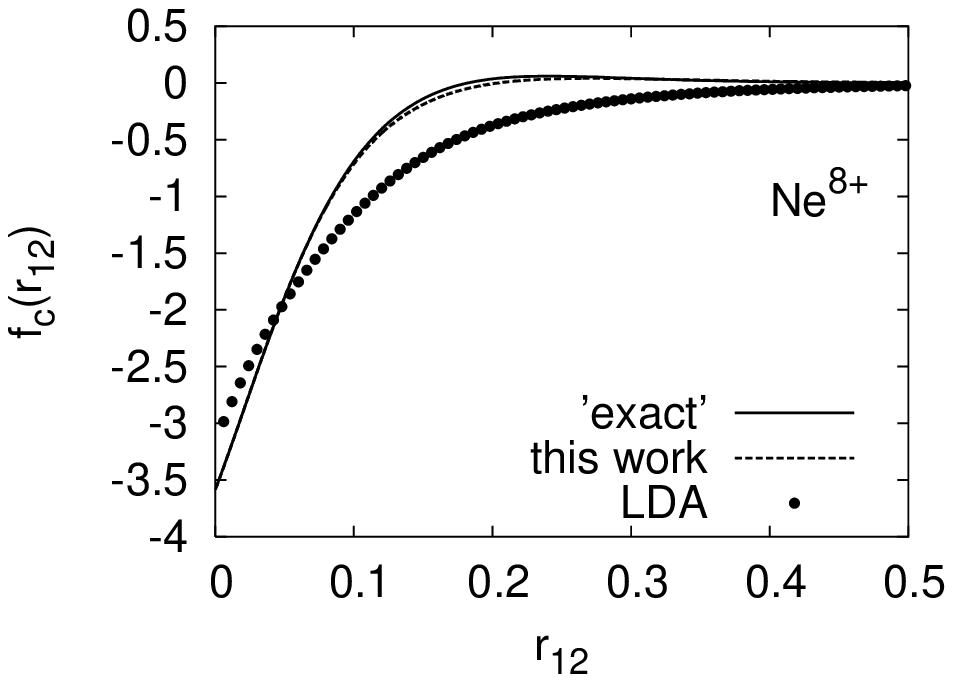} 
\caption{The correlated part of the spherically- and
system-averaged pair density,
$f_c(r_{12})=f(r_{12})-f_{\rm KS}(r_{12})$.
Our results for He and Ne$^{8+}$ are compared
with the ``exact'' ones and with the LDA result (the hole
for the uniform electron gas is taken from Ref.~\cite{GP2}).}
\label{fig_fcLDA}
\end{figure}

\begin{figure}[t]
\includegraphics[width=6.5cm]{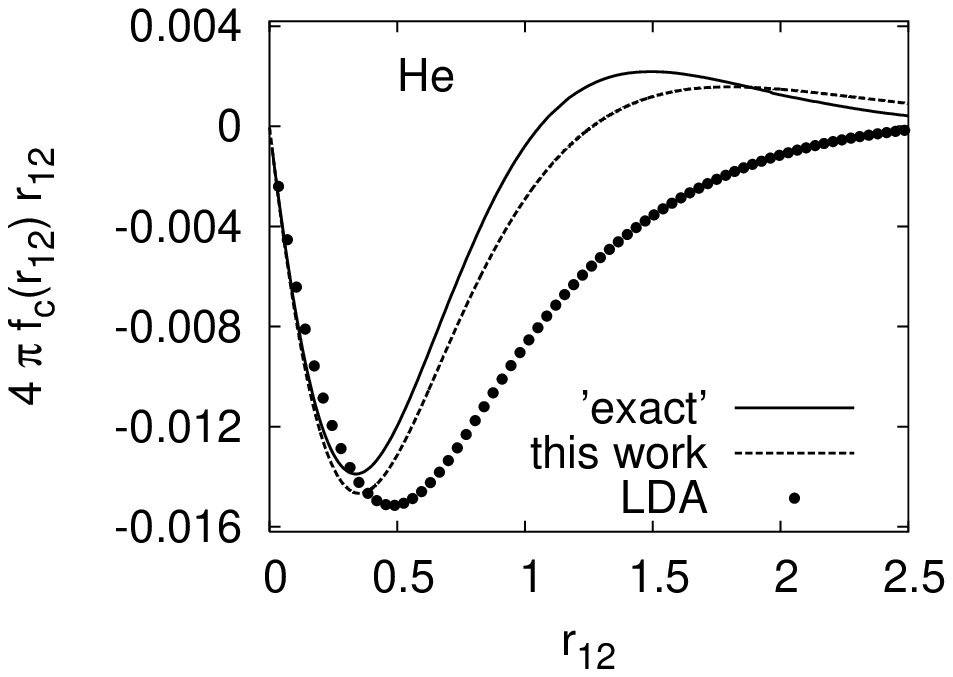} 
\includegraphics[width=6.5cm]{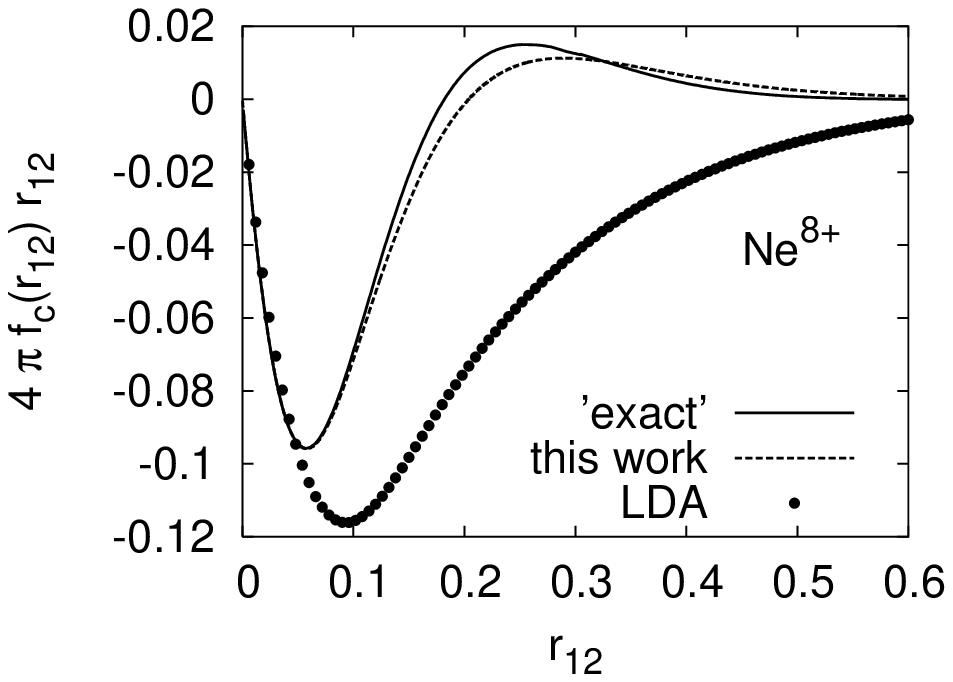} 
\caption{The real space analysis of the correlation
part of the expectation value of $V_{ee}$: the area under each
curve gives $\langle V_{ee}\rangle-\langle V_{ee}\rangle_{KS}$
(see also Eq.~(\ref{eq_Vee}) and Fig.~\ref{fig_fcLDA}).
Our results for He and Ne$^{8+}$ are compared
with the ``exact'' ones and with the LDA result (the hole
for the uniform electron gas is taken from Ref.~\cite{GP2}).}
\label{fig_vcLDA}
\end{figure}

In Fig.~\ref{fig_fcLDA} we consider He and Ne$^{8+}$, and we
compare the correlated
part of our $f$, $f_c=f-f_{\rm KS}$,
with the ``exact'' result~\cite{morgan} and with the corresponding
quantity calculated within LDA, i.e.,
\beq
f_c^{\rm LDA}(r_{12})=\frac{1}{2}\int n(\rv)^2\, g_c(r_{12};
n(\rv))\, d\rv,
\eeq
where $g_c$ is the pair-correlation function of the uniform electron 
gas at full coupling strength, taken from Ref.~\cite{GP2}.
(For an extended system of uniform density $n$, we have $g_c=2 f_c/n N$.)
Figure~\ref{fig_vcLDA} shows the same quantities multiplied
by $4 \pi r_{12}$, i.e. the integrand of Eq.~(\ref{eq_Vee}) for
the correlation part of $\langle V_{ee}\rangle$: the area under
each curve gives $\langle V_{ee}\rangle-\langle V_{ee}\rangle_{KS}$.
In the last line of Table~\ref{tab_resume} we report quantitative
results for $\langle V_{ee}\rangle-\langle V_{ee}\rangle_{KS}$.
This quantity is less accurate than the short-range properties, but
it is still encouraging. Moreover, it saturates for large $Z$ as
in the exact case.

\begin{figure}
\includegraphics[width=6.5cm]{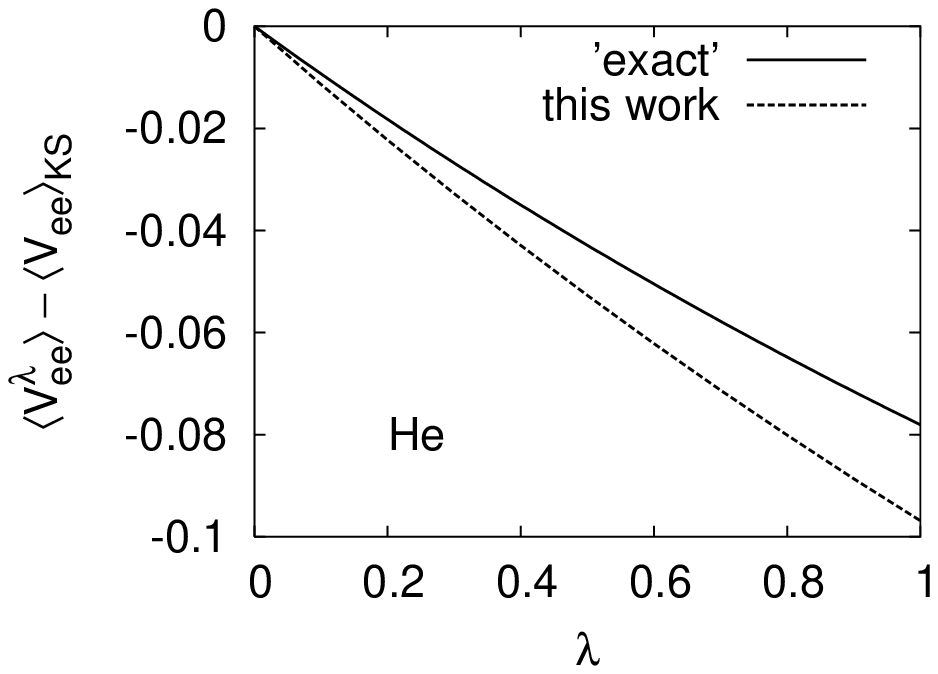}  
\includegraphics[width=6.5cm]{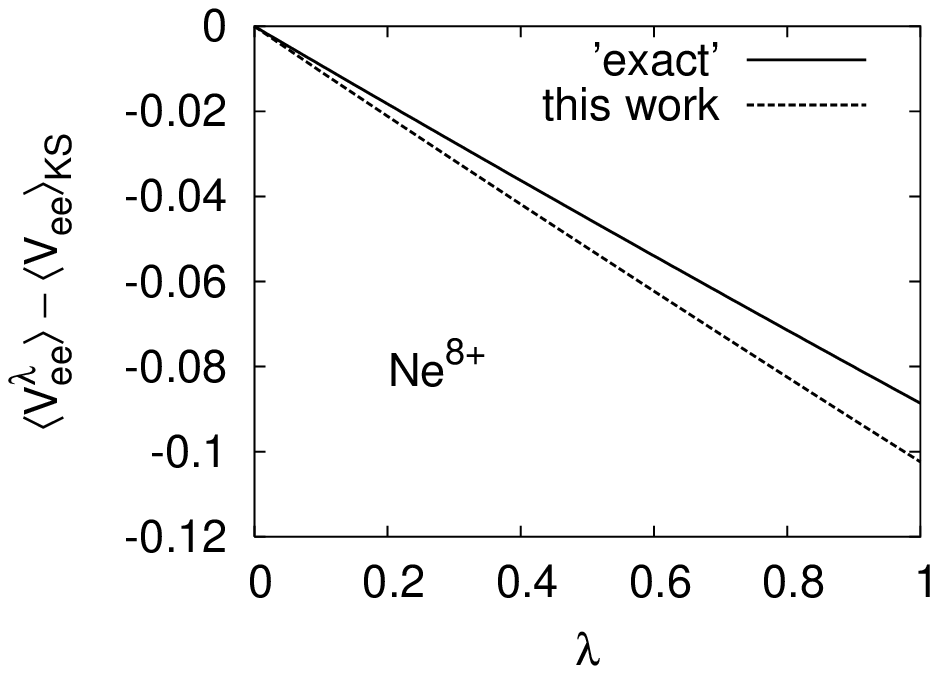} 
\caption{Correlation part of $\langle V_{ee}^{\lambda}\rangle$ 
along the linear 
adiabatic connection for He and Ne$^{8+}$. Our results 
are compared with the ``exact'' ones of Ref.~\cite{francois}. 
The area under each curve gives the correlation energy $E_c$
of standard Kohn-Sham theory.}
\label{fig_lambda}
\end{figure}

\section{Adiabatic connection and correlation energy}
\label{sec_adiabatic}
For the calculation of the energy
of the physical system, in addition to $V_c[n]=\langle V_{ee}\rangle-
\langle V_{ee}\rangle_{KS}$, one needs to know the kinetic-energy difference,
$T_c[n]=\langle T \rangle-\langle T\rangle_{KS}$, that can be obtained
via the adiabatic connection formalism~\cite{adiabatic,gunnarsson,wang}. 
By varying a parameter $\lambda$,
the interaction $v_{ee}^{\lambda}(r_{12})$ 
between the electrons is switched on continuously from
zero to $1/r_{12}$, while the density is kept fixed by an external
one-body potential $V^{\lambda}$. If
$v_{ee}^{\lambda=0}=0$ and $v_{ee}^{\lambda = a}= 1/r_{12}$, the KS
correlation energy $E_c[n]=T_c[n]+V_c[n]$ is given by~\cite{adiabatic,wang}
\beq
E_c[n]=\int_0^a d\lambda \int_0^\infty dr_{12}\, 
4\pi\,r_{12}^2\,f_c^{\lambda}(r_{12})\frac{\partial v_{ee}^{\lambda}(r_{12})}
{\partial \lambda},
\label{eq_adiabatic}
\eeq
where $f_c^{\lambda}=f^{\lambda}-f_{\rm KS}$.  

Usually, the adiabatic connection is performed along a linear
``path''~\cite{gunnarsson,mike}, 
by setting $v_{ee}^{\lambda}=\lambda/r_{12}$, which leads
to Eq.~(\ref{eq_adialinear}). If one is able to compute the exact
$f_c^{\lambda}$, the resulting $E_c$ from Eq.~(\ref{eq_adiabatic}) is
independent of the choice of $v_{ee}^{\lambda}$. However,
when approximations are made some ``paths'' can give much better
results than others~\cite{adiabatic}. As we shall see, this is the case with 
the present approach.

We build an Overhauser-like potential for interaction $v_{ee}^{\lambda}$
(to be added to $v_{\rm eff}^{(0)}$) as
\beq
v_{\rm eff}^{Ov,\,\lambda}(r_{12};\overline{r}_s)=v_{ee}^{\lambda}(r_{12})
-
\int_{|\rv|\le \overline{r}_s} 
\overline{n}\,  v_{ee}^{\lambda}(|\rv - \rv_{12}|)\,d \rv.
\label{eq_vefflambda}
\eeq
That is, the average density $\overline{n}$ of Eq.~(\ref{eq_avn})
(and thus the average $\overline{r}_s$) is kept
fixed to mimic the fact that the one-electron density does not change along the
adiabatic connection. The modified interaction $v_{ee}^{\lambda}$
is screened by a sphere of radius $\overline{r}_s$ and of positive
uniform charge of density $\overline{n}$ that attracts the electrons
with the same modified interaction. This attractive background approximates
the effect of the external potential $V^{\lambda}$ on $f$.

\subsection{Linear adiabatic connection}
If we choose $v_{ee}^{\lambda}=\lambda/r_{12}$ we simply obtain
$v_{\rm eff}^{Ov,\,\lambda}(r_{12};\overline{r}_s)=
\lambda
\,v_{\rm eff}^{Ov}(r_{12};\overline{r}_s)$, where
$v_{\rm eff}^{Ov}(r_{12};\overline{r}_s)$ is given by Eq.~(\ref{eq_potOv}).

The results for $\langle V_{ee}^{\lambda}\rangle -
\langle V_{ee}\rangle_{KS}$ for He and Ne$^{8+}$ are shown
in Fig.~\ref{fig_lambda}, and are compared with the ``exact'' ones
of Ref.~\cite{francois}. The correlation energy $E_c$ can be calculated
as the area under each curve. We obtain $E_c=-0.052$~Hartree for He
and $E_c=-0.053$~Hartree for Ne$^{8+}$, to be compared
with the corresponding 'exact' results, -0.042 and -0.045, respectively.

\subsection{A nonlinear adiabatic connection}
As shown by Figs.~\ref{fig_all}--\ref{fig_vcLDA} 
and Table~\ref{tab_resume}, the
Overhauser-like potential gives accurate results for the short-range
part of $f_c(r_{12})$. We can thus expect to obtain better correlation
energies from the adiabatic connection formalism if we choose
a modified interaction $v_{ee}^{\lambda}$ that is able to separate
long-range and short-range contributions, 
like the ``erf'' interaction~\cite{adiabatic,erf,sav_madrid,julien}
\beq
v_{ee}^{\lambda}(r_{12})=\frac{\erf(\lambda\,r_{12})}{r_{12}}.
\label{eq_erfinte}
\eeq 
With this choice, Eq.~(\ref{eq_adiabatic}) becomes
\beq
E_c[n]=\int_0^\infty d\lambda \int_0^{\infty} dr_{12}\,
 4\pi\,r_{12}^2\,f_c^{\lambda}(r_{12})
\,\tfrac{2}{\sqrt{\pi}}e^{-\lambda^2r_{12}^2}.
\label{eq_adiaerf}
\eeq
For large $\lambda$, when we are approaching 
the physical system, the gaussian factor $e^{-\lambda^2r_{12}^2}$
in Eq.~(\ref{eq_adiaerf}) quenches the long-range contribution 
of $f_c^{\lambda}$ to the
energy integrand. At the KS end of the adiabatic connection,
when $\lambda \to 0$, the interaction, and thus
$f_c^{\lambda}$, become small, so that the contribution to $E_c$
coming from $\lambda$-values for which the long-range part
of $f_c^{\lambda}$ is not quenched is moderate. Moreover,
the function $f_c^{\lambda}$ is correctly normalized to zero
so that for $\lambda\to 0$, not only is $f_c^{\lambda}$ small, but
also the integral itself vanishes. In the linear
adiabatic connection of Eq.~(\ref{eq_adialinear}), instead,
the long-range part of $f_c^{\lambda}$ plays an important
role in the energy integrand at all $\lambda$. 
Indeed,
with this nonlinear adiabatic connection we obtain
$E_c=-0.0405$ Hartree for He and $E_c=-0.0413$ for Ne$^{8+}$, much closer
to the ``exact'' values with respect to the results from the
linear adiabatic connection. 

\begin{figure}
\includegraphics[width=6.5cm]{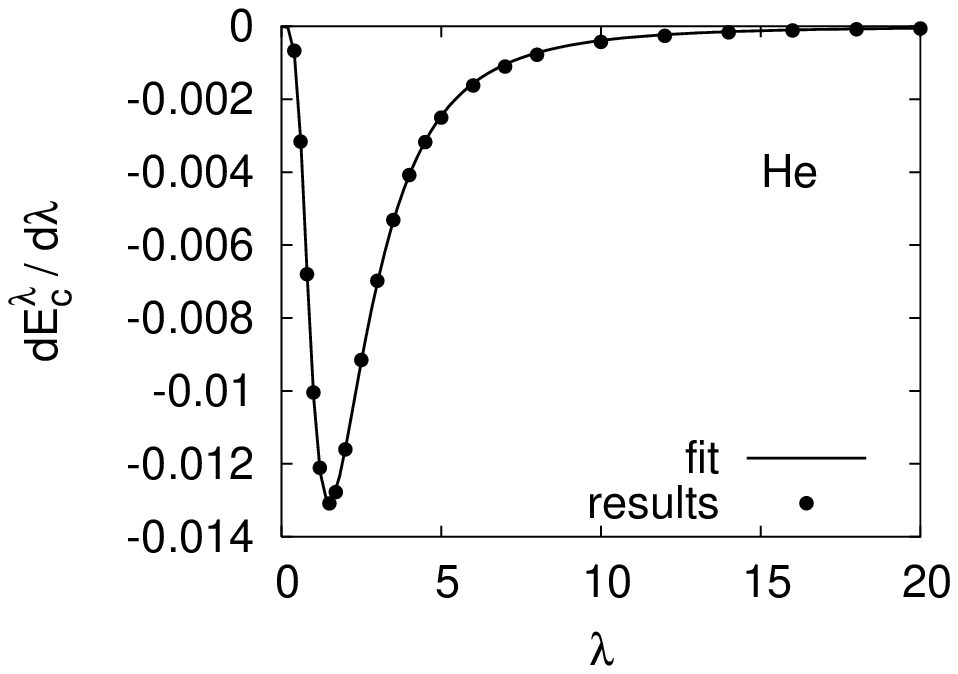}  
\includegraphics[width=6.5cm]{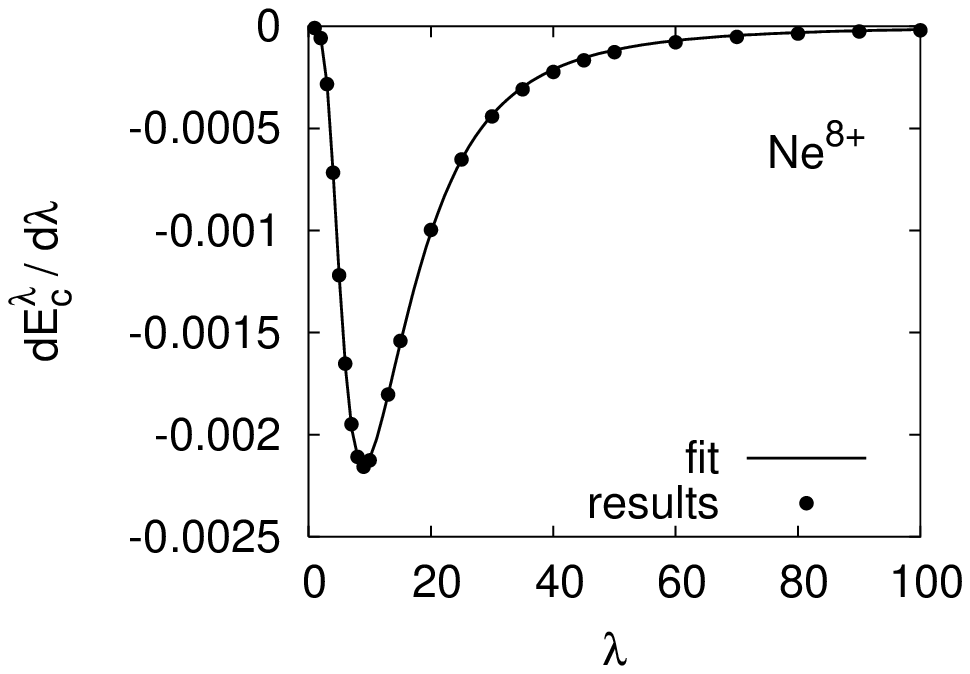} 
\caption{The derivative $dE_c^{\lambda}/d\lambda$ 
along the nonlinear 
adiabatic connection defined by 
Eqs.~(\ref{eq_erfinte})--(\ref{eq_adiaerf})
for He and Ne$^{8+}$. Our results 
are compared with the derivative of the fitting function of
Eq.~(\ref{eq_ecfit}).
The area under each curve from zero to
$\infty$ gives the correlation energy $E_c$
of standard Kohn-Sham theory.}
\label{fig_veeerf}
\end{figure}

The technical details of this calculation are as follows.
The potential $v_{\rm eff}^{Ov,\,\lambda}(r_{12};\overline{r}_s)$
of Eq.~(\ref{eq_vefflambda}) can be computed analytically,
and is reported in Appendix~\ref{app_poterf}. We thus obtained,
via Eqs.~(\ref{eq_eff1})-(\ref{eq_eff2}),
$dE_c^{\lambda}/d\lambda=\int d\rv_{12}f_c^{\lambda}(r_{12})
\,\tfrac{2}{\sqrt{\pi}}e^{-\lambda^2r_{12}^2}$  for 23
values of $\lambda$ between 0 and 20 for He, and
between 0 and 100 for Ne$^{8+}$. We then fitted our results
with the derivative of the following functional form
\beq
E_c^{\lambda}=-\frac{a_1\,x^6+a_2\,x^8+a_3\,x^{10}}
{(1+b^2\,x^2)^5}, \qquad x=\frac{\lambda}{Z},
\label{eq_ecfit}
\eeq
that has exact asymptotic behaviors for small and large 
$\lambda$~\cite{julien}. (We have numerical evidence that
our results fulfill such exact behaviors.) In Fig.~\ref{fig_veeerf}
we report our numerical values for $d E_c^{\lambda}/{d\lambda}$,
together with the derivative of the fitting function of 
Eq.~(\ref{eq_ecfit}). The parameters and the r.m.s
of residuals are reported in Table~\ref{tab_fit}. The KS correlation
energy $E_c[n]$ is then given by $a_3/b^{10}$. 

\begin{figure}
\includegraphics[width=6.5cm]{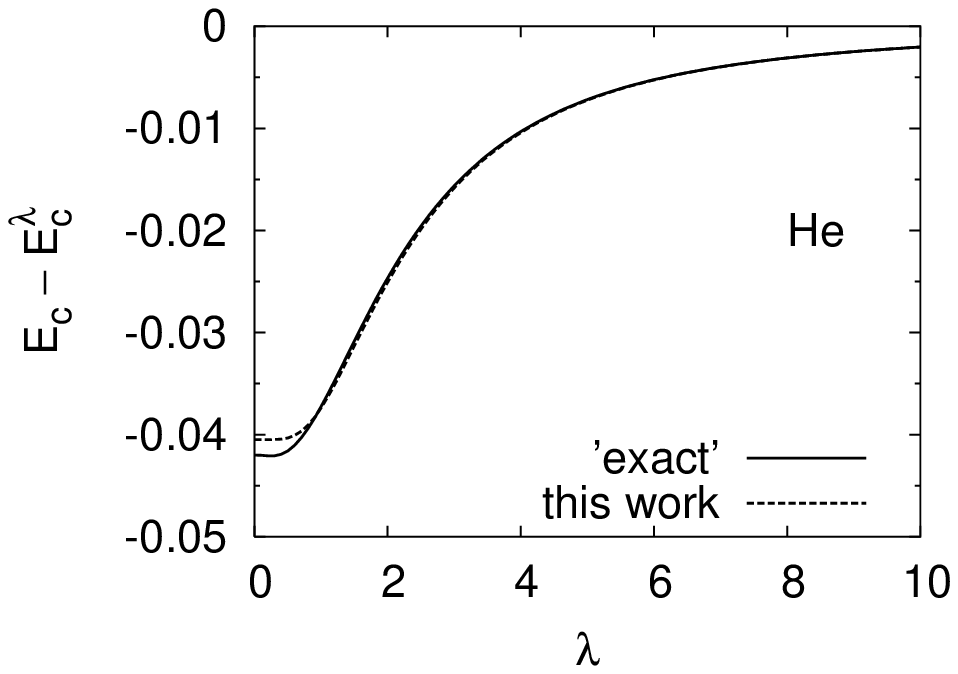}  
\includegraphics[width=6.5cm]{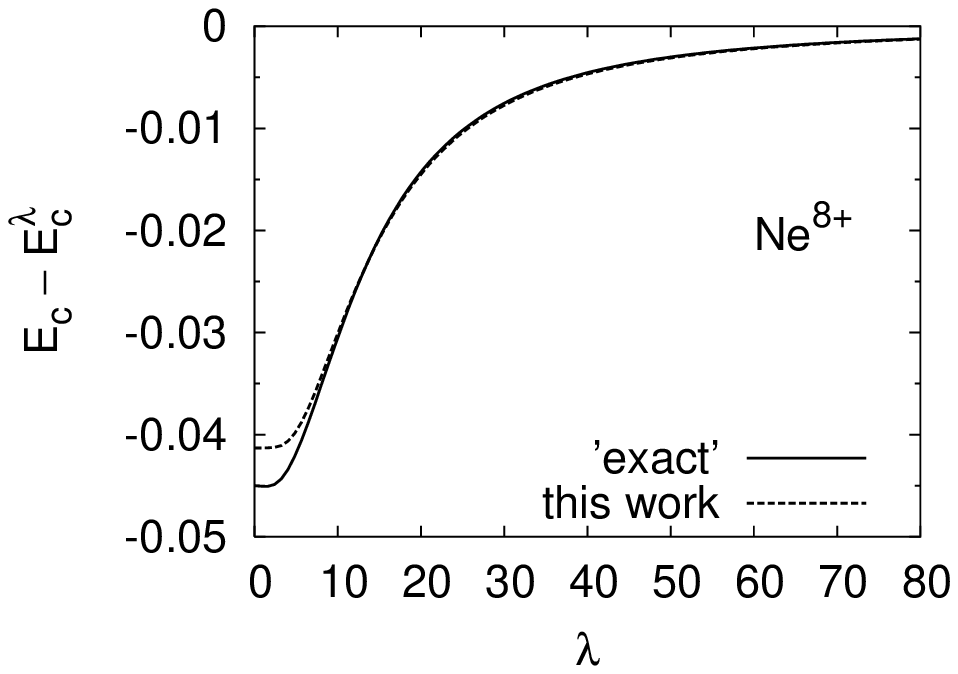} 
\caption{A nonlinear adiabatic connection that
separates long- and short-range effects: 
difference between the correlation energy $E_c$ 
of the physical system (with full interaction $1/r$) and
the correlation energy $E_c^{\lambda}$
of the system with partial interaction $\erf(\lambda\,r)/r$, 
for He and Ne$^{8+}$.
Our results 
are compared with the ``exact'' ones of Ref.~\cite{sav_madrid}.}
\label{fig_ecmu}
\end{figure}

The accuracy of our results with the ``erf'' adiabatic connection
 is of particular interest for the
method of Refs.~\cite{adiabatic,erf,sav_madrid,julien}, 
which combines multideterminantal 
wavefunctions (configuration interaction, CI)
with density functional theory (``CI+DFT''). In such
approach, instead of the KS system, one choses a reference system
of partially interacting particles, usually with the potential of
Eq.~(\ref{eq_erfinte}). This model system is treated with
a multideterminantal wavefunction, in a CI fashion, that allows to treat
near-degeneracy effects. The remaining part of the energy is
calculated via a density functional, that needs to be approximated.
The larger $\lambda$, the larger is the energy fraction treated with the
CI calculation, and thus the larger is the computational cost. 
The correlation energy functional that
needs to be approximated is~\cite{adiabatic,erf,sav_madrid,julien}
\beq
\overline{E}^{\lambda}_c[n]\equiv E_c[n]-E^{\lambda}_c[n],
\eeq
and can be rewritten as
\beq
\overline{E}_c^{\lambda}[n]=\int_\lambda^{\infty} d\lambda' \int_0^{\infty}
dr_{12}\, 4\pi\,r_{12}^2\,f_c^{\lambda'}(r_{12})
\,\tfrac{2}{\sqrt{\pi}}e^{-\lambda'^2r_{12}^2}.
\label{eq_eclambda}
\eeq
Thus, only the short-range part of $f_c^{\lambda}$ contributes 
to the functional $\overline{E}_c^{\lambda}[n]$, and we expect to get
accurate results with the present approach. Indeed, this is the case,
as shown in Fig.~\ref{fig_ecmu}, where we compare our results as
a function of $\lambda$ with the ``exact'' ones of Ref.~\cite{sav_madrid}.
The error is less than 0.5~mHartree for $\lambda \gtrsim
1/\overline{r}_s$, that is a very reasonable choice for the
value of $\lambda$ to be used in the CI+DFT
method of Refs.~\cite{adiabatic,erf,sav_madrid,julien}.

\begin{table}
\begin{tabular}{lccccc}
\hline\hline
 &  $a_1$ & $a_2$ & $a_3$ & $b$ & r.m.s. \\
\hline
He & 1.2047 & 2.3253 & 2.7788 & 1.5263  & 4$\cdot10^{-5}$ \\
& & & & & \\
Ne$^{8+}$ & 0.3983 & 0.4711 & 0.4026  & 1.2557 & 9$\cdot10^{-6}$\\
\hline\hline  
\end{tabular}
\caption{Optimal fit parameters and r.m.s of the residuals for the
derivative of Eq.~(\ref{eq_ecfit}), that parametrizes our results
for $dE_c^{\lambda}/d\lambda$ along the nonlinear 
adiabatic connection defined by 
Eqs.~(\ref{eq_erfinte})--(\ref{eq_adiaerf}). See also Fig.~\ref{fig_veeerf}.}
\label{tab_fit}
\end{table}

\section{Conclusions and perspectives}
\label{sec_last}
In this work
we have started to explore the possibility of solving
simple radial equations to generate realistic spherically-
and system-averaged electronic pair densities $f(r_{12})$
for nonuniform systems. With a simple approximation
for the unknown effective electron-electron interaction that
appears in our formalism, 
we have obtained, for two-electron atoms, results that are in fair
agreement with those coming from accurate variational wavefunctions
(Figs.~\ref{fig_all}--\ref{fig_vcLDA} and Table~\ref{tab_resume}).
We have then extended our approach along a nonlinear
 adiabatic connection and
obtained Kohn-Sham correlation energies whose error is
less than 4~mHartrees, and
short-range-only correlation energies
whose accuracy is one order of magnitude better (Fig.~\ref{fig_ecmu}). 

In Sec.~\ref{sec_formalism}, we have introduced a general formalism
for many-electron systems that will be further tested in future
work. So far we can say that this formalism, combined with simple physical
approximations, works very well for two completely different
systems: the uniform electron 
gas~\cite{GP1,DPAT1,CGPenagy2} and
the He series. We think that this fact makes the method promising.

To fully develop the approach described in this paper, 
many steps have to be performed. 
First of all, the KS part of the effective $e$-$e$ potential,
$v_{\rm eff}^{(0)}(r_{12})$ of Eq.~(\ref{eq_ourv}), should
be also approximated, to make the extension to many-electron systems
practical. The correlation
part of the  effective $e$-$e$ potential can be improved,
in analogy with the recent developments for the uniform electron-gas
case~\cite{DPAT1,CGPenagy2}. It should then be possible to construct
a self-consistent scheme (OEP-like) that combines
the Kohn-Sham equations with the correlation energy functional arising
from our approach [Eqs.~(\ref{eq_eff1})-(\ref{eq_eff2}) at different
coupling strengths $\lambda$, plus 
Eq.~(\ref{eq_adialinear}) or Eq.~(\ref{eq_adiaerf})]. With respect to
traditional DFT calculations, this combined scheme would have the advantage
of yielding not only the ground-state one-electron density
$n(\rv)$ and energy $E$, but also the spherically- and system-averaged
pair density $f(r_{12})$, thus allowing to calculate expectation values
of two-body operators that only depend on the electron-electron distance.
The combination
of our approach with the CI+DFT method
of Refs.~\cite{adiabatic,erf,sav_madrid,julien} could be also implemented
and, in view of the results of Fig.~\ref{fig_ecmu}, it is even more promising.
We are presently working in all of these main directions.

\section*{Acknowledgments}
We thank C. Umrigar for the wavefunctions of Ref.~\cite{morgan},
J. Toulouse for the results of Ref.~\cite{sav_madrid},
E.K.U. Gross, W. Kohn, M. Polini, G. Vignale and P. Ziesche 
for encouraging discussions, V. Sahni
for useful hints, and J.K. Percus for helpful suggestions.
This research was supported by a Marie Curie Intra-European
Fellowships within the 6th European Community Framework 
Programme (contract number MEIF-CT-2003-500026).

\appendix
\section{Overhauser-like potential for the erf interaction}
\label{app_poterf}
The evaluation of Eq.~(\ref{eq_vefflambda}) with the interaction
$v_{ee}^{\lambda}=\erf(\lambda\,r_{12})/r_{12}$ gives
\beq
v_{\rm eff}^{Ov,\,\lambda}(r_{12};\overline{r}_s)=\frac{u(s,\mu)}
{\overline{r}_s},
\eeq
where $s=r_{12}/\overline{r}_s$, $\mu=\lambda \,\overline{r}_s$, and
\begin{widetext}
\begin{eqnarray}
u(s,\mu)  =  \frac{\erf(\mu\,s )}{s} - 
  \frac{1}{8\,{\sqrt{\pi }}\,s\,
     {\mu }^3}\biggl\{2\,\left[ 1 + \left( -2 + s + s^2 \right) 
\,{\mu }^2 + 
     e^{-4\,s\,{\mu }^2} (-1 + \left( 2 + s - s^2 \right) \,{\mu }^2) 
    \right] e^{-{\left( 1 - s \right) }^2\,{\mu }^2} -
\nonumber \\    
 {\sqrt{\pi }}\,\mu  
\left[ 3\,s + 
        2\,{\left( 1 - s \right) }^2\,\left( 2 + s \right) \,{\mu }^2 
\right] \,
      \erf\left[\mu \,\left( 1 - s \right) \right] + 
     {\sqrt{\pi }}\,\mu \,\left[ -3\,s + 
        2\,\left( 2 - s \right) \,{\left( 1 + s \right) }^2\,{\mu }^2 \right] \,
      \erf\left[\mu\,\left( 1 + s \right) \right]\biggr\}.
\end{eqnarray}
\end{widetext}


\begin{thebibliography}{99} 
\bibitem{kohnnobel}
W. Kohn, Rev. Mod. Phys. {\bf 71}, 1253 (1999).
\bibitem{science}
A.E. Mattsson, Science {\bf 298}, 759 (2002).
\bibitem{FNM}
C. Fiolhais, F. Nogueira, and M. Marques (eds.), {\it A Primer in Density
Functional Theory} (Springer-Verlag, Berlin, 2003).
\bibitem{kohnsham}
W. Kohn and L.J. Sham,  Phys. Rev. {\bf 140}, A1133 (1965).
\bibitem{jacob}
J.P. Perdew and K. Schmidt, in {\it Density Functional Theory and Its
Applications to Materials}, edited by V. VanDoren {\it et al.} (AIP, NY, 2001),
and references therein.
\bibitem{adiabatic}
A. Savin, F. Colonna, and R. Pollet, Int. J. Quantum Chem.
{\bf 93}, 166 (2003), and references therein.
\bibitem{sahni} V. Sahni, {\it Quantal Density Functional Theory}
(Springer-Verlag, Berlin, 2004).
\bibitem{GGA}
J.P. Perdew, K. Burke, and M. Ernzerhof, Phys. Rev. Lett. {\bf 77}, 3865
(1996); {\it ibid.} {\bf 78}, 1396 (1997); A.D. Becke, Phys. Rev. A
{\bf 38}, 3098 (1988); J. Chem. Phys. {\bf 84}, 4524 (1986); 
C. Lee, W. Yang, and R.G. Parr, Phys. Rev. B
{\bf 37}, 785 (1988); J.P. Perdew, J.A. Chevary, S.H. Vosko, K.A. Jackson,
M.R. Pederson, D.J. Singh, and C. Fiolhais, Phys. Rev. B. {\bf 46}, 6671
(1992); {\it ibid.} {\bf 48}, 4978 (1993).
\bibitem{TPSS}
J. Tao, J.P. Perdew, V.N. Staroverov, and G.E. Scuseria,
Phys. Rev. Lett. {\bf 91}, 146401 (2003).
\bibitem{newjohn}
J.P. Perdew, A. Ruzsinszky, J. Tao, V.N. Staroverov, G.E. Scuseria, and
G.I. Csonka, J. Chem. Phys., to appear.
\bibitem{mike}
M. Seidl, J.P. Perdew, and S. Kurth, Phys. Rev. Lett. {\bf 84}, 5070 (2000).
\bibitem{oep}
see, e.g.,
S. K\"ummel and J.P. Perdew, Phys. Rev. B {\bf 68}, 035103 (2003);
W. Yang and Q. Wu, Phys. Rev. Lett. {\bf 89}, 143002 (2002);
R. J. Magyar, A. Fleszar, and E. K. U. Gross,
Phys. Rev. B {\bf 69}, 045111 (2004);
M. Gr\"uning, O. V. Gritsenko, and E. J. Baerends,
J. Chem. Phys. {\bf 118}, 7183 (2003). 
\bibitem{Coulson}
C.A. Coulson and A.H. Neilson, Proc. Phys. Soc. London {\bf 78},
831 (1961).
\bibitem{Cioslowski1}
J. Cioslowski, B.B. Stefanov, A. Tan, and C.J. Umrigar,
J. Chem. Phys. {\bf 103}, 6093 (1995).
\bibitem{Cioslowski2}
J. Cioslowski and G. Liu, J. Chem. Phys. {\bf 109}, 8225 (1998).
\bibitem{ugalde1}
E. Valderrama, J.M. Ugalde, and R.J. Boyd, in {\it Many-electron
densities and reduced density matrices}, edited by J. Cioslowski
(Kluwer Academic/Plenum Publishers, New York, 2000).
\bibitem{davidson}
E.R. Davidson, {\it Reduced Density Matrices in Quantum Chemistry}
(Academic Press, New York, 1976).
\bibitem{coleman}
A.J. Coleman and V.I. Yukalov, {\it Reduced Density Matrices:
Coulson's Challenge} (Springer-Verlag, New York, 2000).
\bibitem{sacchetti} see, e.g., G. Mazzone, F. Sacchetti, and V. Contini,
Phys. Rev. B {\bf 28}, 1772 (1983);
C. Petrillo and F. Sacchetti, {\it ibid.} {\bf 51}, 4755 (1995).
\bibitem{cusp} A. K. Rajagopal, J. C. Kimball, and M. Banerjee,
        Phys. Rev. B {\bf 18}, 2339 (1978);
X.-Y. Pan and V. Sahni, J. Chem. Phys. {\bf 119}, 7083 (2003).
\bibitem{nota}
The difference in kinetic energy can be obtained from the
adiabatic connection formula that can be viewed as a generalization of
the above argument, and is treated in Sec.~\ref{sec_adiabatic}.
\bibitem{Ov}
A.W. Overhauser, Can. J. Phys. {\bf 73}, 683 (1995).
\bibitem{GP1} P. Gori-Giorgi and J.P. Perdew, Phys. Rev. B
{\bf 64}, 155102 (2001).
\bibitem{DPAT1} B. Davoudi, M. Polini, R. Asgari, and M.P. Tosi,
Phys. Rev. B {\bf 66}, 075110 (2002).
\bibitem{CGPenagy2} M. Corona, P. Gori-Giorgi, and J.P. Perdew,
Phys. Rev. B {\bf 69}, 045108 (2004);
I. Nagy, R. Diez Mui\~no, J.I. Juaristi, and P.M. Echenique,
Phys. Rev. B {\bf 69}, 233105 (2004).
\bibitem{ziesche1}
P. Ziesche, Phys. Lett. A {\bf 195}, 213 (1994);
M. Levy and P. Ziesche, J. Chem. Phys. {\bf 115}, 9110 (2001).
\bibitem{gonis}
A. Gonis, T.G. Schulthess, J. van Ek, and P.E.A. Turchi,
Phys. Rev. Lett. {\bf 77}, 2981 (1996); A. Gonis, T.G. Schulthess, P.E.A. 
Turchi, and J. van Ek, Phys. Rev. B {\bf 56}, 9335 (1997).
\bibitem{agnesnagy} 
A. Nagy, Phys. Rev. A {\bf 66}, 022505 (2002).
\bibitem{furche} F. Furche, Phys. Rev. A {\bf 70}, 022514 (2004).
\bibitem{soler}
J.M. Soler, Phys. Rev. B {\bf 69}, 195101 (2004).
\bibitem{levy}
M. Levy, Proc. Natl. Acad. Sci. U.S.A. {\bf 76}, 6062 (1979).
\bibitem{lieb} E. Lieb, Int. J. Quantum Chem. 
{\bf 24}, 243 (1983). 
\bibitem{gunnarsson}
J. Harris and R. Jones, J. Phys. F {\bf 4}, 1170 (1974);
D.C. Langreth and J.P. Perdew, Solid State Commun. {\bf 17}, 1425 (1975);
O. Gunnarsson and B.I. Lundqvist, Phys. Rev. B {\bf 13}, 4274 (1976). 
\bibitem{wang}
W. Yang, J. Chem. Phys. {\bf 109}, 10107 (1998).
\bibitem{erf}
A. Savin, in {\it Recent Developments and Applications of Modern 
Density Functional Theory}, edited by J.M. Seminario 
(Elsevier, Amsterdam, 1996); T. Leininger, H. Stoll, H.-J. Werner, and 
A. Savin, Chem. Phys. Lett. {\bf 275}, 151 (1997);
R. Pollet, A. Savin, T. Leininger, and H. Stoll,
J. Chem. Phys. {\bf 116}, 1250 (2002).
\bibitem{sav_madrid}
R. Pollet, F. Colonna, T. Leininger, H. Stoll, H.-J. Werner, and A. Savin,
Int. J. Quantum Chem. {\bf 91}, 84 (2003); J. Toulouse, private
communication.
\bibitem{julien} J. Toulouse, F. Colonna, and A. Savin, Phys. Rev. A 
{\bf 70}, 062505 (2004). 
\bibitem{tosi2}
B. Davoudi, R. Asgari, M. Polini, and M.P. Tosi, Phys. Rev. B
{\bf 68}, 155112 (2003); R. Asgari, B. Davoudi, and M.P. Tosi,
Solid State Comm. {\bf 131}, 301 (2004).
\bibitem{pisani_new}
R. Asgari, B. Davoudi, M. Polini, and M.P. Tosi, in preparation.
\bibitem{morgan}
D.E. Freund, B.D. Huxtable, and J.D. Morgan III,
Phys. Rev. A {\bf 29}, 980 (1984). We used an improved version
(provided to us by C. Umrigar) of the
accurate variational wavefunctions
described in this work to obtain 
one-electron densities $n(\rv)$ and functions $f(r_{12})$. See also
C.J. Umrigar and X. Gonze, Phys. Rev. A {\bf 50}, 3827 (1994), and
Ref.~\cite{Cioslowski1}.
\bibitem{ziesche2}
P. Ziesche, Phys. Rev. B {\bf 67}, 233102 (2003);
P. Ziesche, K. Pernal, and F. Tasn\'adi, 
Phys. Status Solidi B {\bf 239}, 185 (2003). In these papers interesting
exact properties
of the ``Overhauser geminals'' are derived, using the
{\em interacting} momentum distribution to define the geminal occupancy.
Notice however that the uniform-electron gas 
equivalent of our Eqs.~(\ref{eq_eff1})-(\ref{eq_eff2}) employ
the {\em non-interacting} momentum distribution, as it has been done in
Refs.~\cite{GP1,DPAT1,CGPenagy2}. 
\bibitem{francois} F. Colonna and A. Savin,
J. Chem. Phys. {\bf 110}, 2828 (1999).
\bibitem{parr}
Q. Zhao, R. C. Morrison, and R. G. Parr,  Phys. Rev. A {\bf 50}, 2138 (1994);
R. van Leeuwen and E. J. Baerends Phys. Rev. A {\bf 49}, 2421 (1994).
\bibitem{ontop}
 J.P. Perdew, A. Savin, and K. Burke, 
Phys. Rev. A, {\bf 51}, 4531 (1995);
K. Burke, J.P. Perdew, and M. Ernzerhof,
J. Chem. Phys. {\bf 109}, 3760 (1998); E. Valderrama and
J.M. Ugalde, Int. J. Quantum Chem. {\bf 86}, 40 (2002). 
\bibitem{ugalde2000}
X. Fradera, M. Duran, E. Valderrama, and J.M. Ugalde,
Phys. Rev. A {\bf 62}, 034502 (2000).
\bibitem{GP2} P. Gori-Giorgi and J.P. Perdew, Phys. Rev.
B {\bf 66}, 165118 (2002).
\end{thebibliography}
\end{document}